\documentclass[aps,10pt,twocolumn,nofootinbib,floatfix,prl]{revtex4}
\usepackage{amsmath}
\usepackage{amsfonts}
\usepackage{amsfonts}
\usepackage{mathrsfs}
\usepackage[mathscr]{euscript}
\usepackage[dvips]{graphicx}
\usepackage{fancyhdr}
\usepackage{colordvi}
\usepackage{soul}

\newcommand{\RealPart}{\mathbb{R}\mathrm{e}}
\newcommand{\ImagPart}{\mathbb{I}\mathrm{m}}

\newcommand{\Tanh}{\mathrm{tanh}}

\begin{document}
\title{Causality-based criteria for a negative refractive index
 must be used with care}
\author{P. Kinsler}
\email{Dr.Paul.Kinsler@physics.org}
\author{M. W. McCall}
\email{m.mccall@imperial.ac.uk}
\affiliation{
  Blackett Laboratory, Imperial College,
  Prince Consort Road,
  London SW7 2AZ,
  United Kingdom.
}

\lhead{\includegraphics[height=5mm,angle=0]{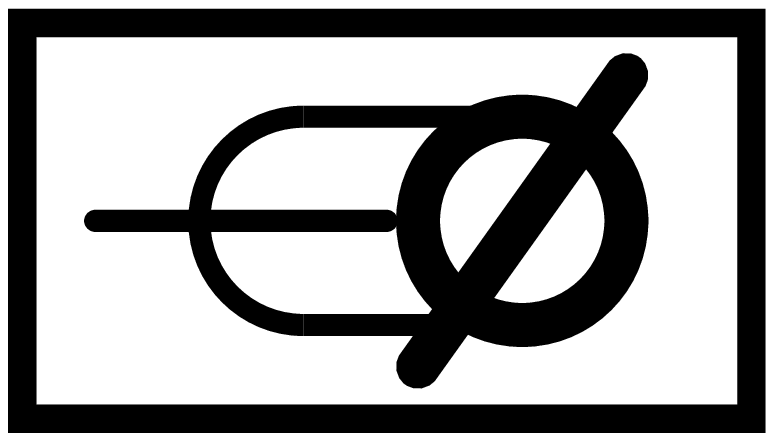}~~MCRIT}
\chead{Causality-based criteria for negative refractive index ...}
\rhead{
Dr.Paul.Kinsler@physics.org\\
http://www.kinsler.org/physics/
}

\begin{abstract}

Using the principle of causality as expressed
 in the Kramers-Kronig relations,
 we derive a generalized criterion for a negative refractive index
 that admits imperfect transparency
 at an observation frequency $\omega$.
It also allows us to relate the global properties of the loss 
 (i.e. its frequency response)
 to its local behaviour at $\omega$.
However, 
 causality-based criteria rely the on the group velocity, 
 not the Poynting vector.
Since the two are not equivalent, 
 we provide some simple examples to compare the two criteria.

\end{abstract}

\date{\today}
\maketitle
\thispagestyle{fancy}

\emph{This is a slightly longer version of the paper published 
 as \textrm{Phys. Rev. Lett. \textbf{101}, 167401 (2008)}; 
 the extra material is largely that removed to fit within the
 4-page length limit, although fig.3 was removed as superfluous 
 during review.}

%

Remarkable electromagnetic properties can be seen  
 in materials engineered so that the phase velocity
 of electromagnetic-wave propagation
 opposes the electromagnetic power flow; 
 such materials are often called ``left handed''
 (see e.g. 
 \cite{Veselago-1968spu,Pendry-2003oe,Smith-K-2000prl,Shelby-SS-2001s,
  Yen-PFVSPBZ-2004sci,Linden-EWZKS-2004sci,Pendry-SS-2006sci,
  Dolling-EWSL-2006sci,Leonhardt-2006sci}), 
 but are more precisely described as
 negative phase velocity (NPV) media.
As might be expected in a rapidly evolving field of research,
 a variety of conditions
 \cite{McCall-LW-2002ejp,Pokrovsky-E-2002ssc,Depine-L-2004motl}
 for NPV have been
 proposed in the literature.
The presence of a negative refractive index (NRI)
 allows for a number of intriguing possibilities,
 e.g.: 
 the creation of a ``perfect lens'' that produces an undistorted image
 without causing any surface reflections \cite{Pendry-2000prl}, 
 the possibility of a reversed Casimir force 
 being used to levitate ultrathin mirrors
 \cite{Leonhardt-P-2007njp}, 
 the automatic compensation of dissipation or dispersion 
 to enhance quantum interference 
 \cite{Yang-XCZ-2008prl}, 
 or the possibility of ``trapped rainbow'' light storage
 \cite{Tsakmakidis-BH-2007n}.


The dispersive nature of the effective medium parameters 
 is exploited in metamaterials to produce an NRI, 
 as confirmed through experimental, 
 theoretical, 
 and numerical studies 
 \cite{Pendry-2003oe,Smith-K-2000prl,Shelby-SS-2001s,Yen-PFVSPBZ-2004sci,
 Linden-EWZKS-2004sci,Pendry-SS-2006sci,Dolling-EWSL-2006sci,
 Leonhardt-2006sci,Wongkasem-ALTKG-2006pier}).
Such metamaterials therefore inherit unavoidable losses
 on the grounds of causality.
Since losses can cause a significant drop in performance, 
 a key challenge is to successfully compensate 
 for loss by adding gain; 
 but note that care must be taken in theoretical investigations
 to ensure that the gain model 
 is both stable and causal \cite{Skaar-2006pre,Boardman-RKM-2007josab}; 
 we also need to use the correct NPV criterion \cite{Kinsler-M-2008motl}.
Here we specifically address the role of the losses required by causality,
 by considering the famous Kramers-Kronig (KK) relations
 (see e.g. \cite{LandauLifshitz})
 which control the relationship between the real and imaginary parts
 of the electric and magnetic material responses
 (i.e. the permittivity $\epsilon$ and permeability $\mu$).
Such relations can also be established for the square
 of the refractive index $n^2 = c^2 \epsilon \mu$, 
 as this quantity inherits the analytical properties
 of $\epsilon(\omega)$ and $\mu(\omega)$: 
 i.e. it lacks singularities in the upper half-plane of complex $\omega$, 
 and $n^2(\omega) \rightarrow 1$ as $\omega \rightarrow \infty$
 (see e.g. \cite{LandauLifshitz,Skaar-2006pre}).
In a recent Letter, 
 Stockman \cite{Stockman-2007prl} adapted the KK relation on $n^2$ 
 to place limits on the \emph{minimum} losses
 that accompany NRI
 for a medium which is perfectly transparent
 at the observation frequency.
He concluded that any significant reduction in the
 losses near the chosen observation frequency will
 also eliminate the NRI.
Whether real metamaterials can 
 \emph{in principle} be made with low loss
 is a question of utmost importance
 in practical metamaterial design. 
Previous work \cite{Stockman-2007prl}
 claimed that the answer is emphatically negative, 
 but we show here that the answer is actually affirmative.

%

In this Letter we replace Stockman's zero-loss criterion
 with another causality-based criterion, 
 one capable of giving useful answers for NPV propagation
 because it admits arbitrary linear optical losses
 both at and away from the observation frequency.
Here we assume a homogeneous medium 
 with $\epsilon$ and $\mu$ being effective parameters obtained 
 for the composite metamaterial
 by (e.g.) a modified S-parameter technique 
 \cite{Smith-VKS-2005pre,Starr-RSN-2004prb}.
Such effective medium approaches 
 are less reliable in the short wavelength 
 (high frequency) regime; 
 but existing analytic attempts only apply to (at best)
 thin composite layers \cite{Saenz-IGT-2007jap}.
The KK relation for $n^2(\omega)$ 
 can be written
~
\begin{align}
  \RealPart \left( n^2 \right) - 1
&=
  \frac{2}{\pi}
  \mathscr{P}
  \int_0^{\infty}
    \frac{\ImagPart (n^2)}
         {s^2 - \omega^2}
      s ds
    \label{eqn-KK-n2guess-R}
,
\end{align}
where $\RealPart()$ and $\ImagPart()$ take the real and imaginary parts;
 thus $\ImagPart \left( n^2 \right) = \epsilon' \mu'' + \epsilon''\mu'$, 
 where $\epsilon'$ and $\mu'$ are the real parts
 of $\epsilon$ and $\mu$, 
 with the imaginary parts being $\epsilon''$ and $\mu''$;
 $\mathscr{P}$ takes the Cauchy principal value.
Noting that the loss in the material is important in the calculation 
 of $\ImagPart \left( n^2 \right)$, 
 Stockman transformed Eq.~(\ref{eqn-KK-n2guess-R}) into one 
 relating the material loss 
 to the presence of NRI,
~
\begin{align}
  \frac{c^2}{v_{p} v_{g}}
&=
  1 + \frac{2}{\pi}
  \int_0^\infty
      \frac{ \ImagPart \left( n^2(s) \right) }
           {\left(s^2 - \omega^2\right)^2}
     s^3 ~
    ds
<0
,
\label{eqn-cf-S3}
\end{align}
by applying the operation 
 $\mathscr{L} = \omega^{-1} \partial_\omega \omega^2$ to both sides
 of Eq.~(\ref{eqn-KK-n2guess-R}).
The behavior of several experimental systems was claimed in
 \cite{Stockman-2007prl} to be consistent with this criterion.
Here the NRI condition 
 relies on opposed (real valued) phase and group velocities, 
 i.e. $v_{p} v_{g} < 0$.
Since 
 the usual acronym NPV is ambiguous,
 we refer to $v_{p} v_{g} < 0$ 
 as NPVG (i.e. NPV w.r.t. group velocity);
 the usual case\footnote{
  The usual NPV condition is $\vec{P} \cdot \vec{k}<0$, 
  with Poynting vector $\vec{P}$ and wavevector $\vec{k}$.
  Using the electromagnetic energy density $\rho$, 
  we can define an energy velocity 
  $v_{E}=\vec{P} \cdot \vec{k}/\rho |\vec{k}|$.
  This allows us to re-express $\vec{P}.\vec{k}<0$ as $v_{p} v_{E} <0$.
  The definition can also be extended to embrace moving media, 
  where $\vec{P}$ is replaced by the 
  electromagnetic energy momentum tensor\cite{McCall-2008meta}.}
 is then NPVE (i.e. NPV w.r.t. energy velocity).

When $v_{p} v_{g} < 0$,
 the integral on the right-hand side (RHS)
 of Eq.~(\ref{eqn-cf-S3}) must be negative.
Stockman therefore concluded that
 even if the losses vanish
 at the observation frequency,
 there must still be significant loss nearby,
 otherwise the integral will produce a positive result.
Consequently,
 systems with imperfect tuning
 or an insufficiently narrow operating bandwidth
 would have their performance degraded.

The limitations of Eq. (\ref{eqn-cf-S3}) are threefold:
\begin{enumerate}

\item
 $\ImagPart (n^2)$ and its derivative
 must be exactly zero at the observation frequency $\omega$ --
 otherwise the integral diverges,
 and the constraint becomes uncertain.

\item
 It only applies at a particular observation frequency, 
 despite utilizing the global properties of the material response.
 It can be used to infer the presence of nearby loss, 
 but does not indicate whether NRI is present there.

\item
 The NPVG condition $v_{p} v_{g} < 0$ is not equivalent
 to the NPVE condition $\vec{P} \cdot \vec{k} < 0$.

\end{enumerate}

These limitations 
 make it hard to determine how losses might be minimized 
 whilst still maintaining NPV over some frequency window.

%

We now replace Stockman's criterion
 with one that avoids divergences 
 while allowing for non-zero loss; 
 thus removing the first two limitations given above.
All necessary convergence or limiting properties 
 for $n^2$ can be satisfied
 if $\epsilon$ and $\mu$ are described by functions of $s$
 which are both rational and causal\footnote{If necessary, 
  inconvenient singularities or divergences in $n^2(\omega)$
  can be removed by considering $f(\omega)n^2(\omega)$, 
  where $f$ is some rational function of $\omega$ designed
  to cancel the pole or to remove the divergence\cite{Toll-1956pr}.}.
The third limitation
 is intrinsic to the approach,
 but has the advantage that it also enables us
 to evaluate the presence of NRI 
 (or, strictly, the presence of NPVG)
 using causality.
Here the group velocity $v_g$ amounts to the commonly used 
 $\partial_\omega k(\omega)$; 
 although imperfect in the case of loss or gain
 (see e.g. \cite{Censor-1977jpa})
 it is that which follows most naturally here.

Our first step is to integrate the RHS
 of Eq.~(\ref{eqn-KK-n2guess-R}) by parts, 
 but only after splitting it into two pieces
 covering the ranges $[0,\omega-\sigma)$ and
 $(\omega+\sigma,\infty)$, 
 then taking the limit $\sigma \rightarrow 0$ at the end.
After defining 
 ${Q}_j = \partial_s^j \ImagPart \left( n^2(s) \right)$, 
and with ${Q}_0(s)= \ImagPart \left( n^2 (s) \right)$ 
 tending to zero fast enough 
 so that the $s=\infty$ surface term vanishes, 
 we find
~
\begin{align}
   \int_{0}^{\infty}
        \frac{{Q}_0(s)}
             {s^2 - \omega^2}
        s ~
        ds
&= 
  -  
  \int_{0}^{\infty}
     \frac{
       {Q}_1(s)
     }{2}
    \ln \left| 1 - \frac{s^2}{\omega^2} \right|
    ~ ds
. \label{eqn-stockman-altB-integral}
\end{align}
Since the RHS is independent of ${Q}_0$, 
 we can now obtain a criterion 
 valid where loss is present
 at the observation frequency; 
 and the better behaved integrand
 means it is considerably easier to make inferences
 about the presence of NRI.
After applying $\mathscr{L}$, 
 we find that 
 $v_{p} v_{g} < 0$ requires 
~
\begin{align}
  \pi
&\le
    \int_{0}^{\infty}
       {Q}_1(s)
    \left[
      \ln \left| 1 - \frac{s^2}{\omega^2} \right|
     +
      \frac{s^2}
           {\omega^2 - s^2 }
      \right]
     ~ ds
.
\end{align}

Before taking $\sigma \rightarrow 0$, 
 a second integration by parts yields
~
\begin{align}
  \pi
&\le
    \int_{0}^{\infty}
       {Q}_1(s)
      \ln \left| 1 - \frac{s^2}{\omega^2} \right|
     ~ ds
\nonumber
\\
&
 +
    \omega
  \int_0^{\omega-\sigma}
       {Q}_2(s)
    \left[
      \Tanh^{-1} \left(\frac{s}{\omega}\right)
     -
      \frac{s}{\omega}
    \right]
    ~ ds
\nonumber
\\
& +
    \omega
  \int_{\omega+\sigma}^{\infty}
       {Q}_2(s)
    \left[
      \Tanh^{-1} \left(\frac{\omega}{s}\right)
     -
      \frac{s}{\omega}
    \right]
    ~ ds
.
\end{align}

Again the surface terms at
 $\omega-\sigma$ and $\omega+\sigma$ will cancel; 
 those at $0$ and $\infty$ vanish. 
With $z=s/\omega$,
 we now have
~
\begin{align}
  \pi
&\le
  \omega
    \int_{0}^{\infty}
      {Q}_1 (z\omega)
      \ln \left| 1 - z^2 \right|
     ~ dz
\nonumber
\\
& -
    \omega^2
  \int_{0}^{1}
    \left[
      {Q}_2 (\omega z)
     +
      \frac{1}{z^2}
      {Q}_2 (\frac{\omega}{z})
    \right]
    \Tanh^{-1} (z)
    ~ dz
.
\label{eqn-integralRHS}
\end{align}
The first (${Q}_1$) term can again be integrated by parts, 
 however the result adds little insight.

Eq.~(\ref{eqn-integralRHS}) is the most general 
 causality-based criterion achievable,
 and is, 
 crucially, 
 not restricted to points of perfect transparency.
It depends only on how the loss 
 (as specified by ${Q}_0 = \ImagPart(n^2)$)
 changes with frequency
 (i.e. on its dispersion, as given by ${Q}_1$ and ${Q}_2$), 
 and not on its magnitude.
Notably,
 the sign of $Q_1$ 
 (i.e. whether $Q_0$
  is increasing or decreasing with frequency)
 has a strong effect on the presence of NRI;
 as does the sign of $Q_2$
 (crudely, 
  whether $Q_0$ is near a minimum or maximum).

The non-${Q}_i$ parts of the integrands
 (i.e. $\ln |1-z^2|$ and $\Tanh^{-1}(z)$)
 are both strongly peaked at $s=\omega$, 
 but nevertheless have finite integrals.
Using the expansion
 $\ImagPart(n^2(s))
 \simeq
  {Q}_0(\omega)
 +
  x {Q}_1(\omega) + (x^2/2) {Q}_2(\omega)$, 
 for $x=s/\omega-1$, 
 we can integrate Eq.~(\ref{eqn-integralRHS}) analytically 
 in an attempt to obtain an approximate criterion
~
\begin{align}
  \pi
&\lesssim
 -
  1.344 ~
  \omega
  {Q}_1(\omega)
    \left[
      1 - 2 \omega {Q}_2 (\omega)
    \right]
 -
  1.386 ~
  \omega^2
  {Q}_2 (\omega)
.
\label{eqn-integralRHSapprox}
\end{align}
Unfortunately
 this fails to convincingly match Eq.~(\ref{eqn-integralRHS}), 
 and the attempt only succeeds in
 emphasizing that it is the \emph{global} properties of the loss
 which constrain the presence of NRI.
Only in Stockman's zero-loss case 
 is a simple intuition valid.

Using our causality-based criterion in Eq.~(\ref{eqn-integralRHS}),
 we can now try to infer
 whether the global properties of $\ImagPart (n^2)$ promote
 (or hinder) NPVG.
Since $v_{p} v_{g}$ and Eq.~(\ref{eqn-integralRHS})
 are intimately connected by the KK relations for $n^2$,
 we used this to numerically test the examples below; 
 nevertheless, 
 each expression provides its own unique perspective --
 one local, 
 one global.
Since we may not always be able to rely
 on obtaining $n^2$ from a model 
 (as in \cite{Smith-VKS-2005pre,Starr-RSN-2004prb}),
 we may need to recover it
 from experimental data.
Whilst the standard KK relations
 are prone to generating inaccurate reconstructions, 
 approaches such as the Multiply Subtractive KK
 method can resolve this for many practical applications --
 even nonlinear spectroscopy \cite{Peiponen-LSV-2004as}.


%

%

We now proceed to test our causality-based criterion.
Since we wish to emphasize general principles,
 and ensure the points we make are clear,
 we consider simple examples with $\epsilon = \mu$,
 rather than more complicated systems;
 we also normalize with respect to
 some suitable reference frequency.
The condition natural to the approach used here is the 
 NPVG one 
 (i.e. $\chi_{G} = v_{p} v_{g} <0$).
This means  
 we only need to calculate 
 (and show) one of $\chi_{G}$ or Eq.~(\ref{eqn-integralRHS});
 we label the result $\chi_{G}$.
In contrast, 
 the NPVE condition
 requires that the phase velocity is opposed to the energy velocity.
This occurs if \cite{Depine-L-2004motl,Kinsler-M-2008motl}
~
\begin{align}
  \chi_{E}
&=
  \epsilon'\left|\mu\right| + \mu'\left|\epsilon\right| < 0
.
\end{align}

The two conditions
 ($\chi_{G}<0$ and $\chi_{E}<0$) will agree
 if the group velocity $v_{g}$
 and energy velocity $v_{E}$ have the same sign.
However
 this only hold in the limit of
 nearly undistorted pulse propagation
 \cite{Milonni-FLSLLHL},
 i.e. for small dispersion and loss.
This is likely to be a poor approximation in NRI materials,
 which by their nature rely on strong dielectric or magnetic response.
So although our criterion in Eq.~(\ref{eqn-integralRHS})
 can be always used to judge the presence of NPVG,
 and make inferences thereon,
 this is not strictly equivalent to the presence of NPVE.

%

Our first example is
 a simple double-plasmon resonance,
 as in e.g. \cite{Smith-SP-2002apl}, setting $\epsilon$ and $\mu$
 according to
~
\begin{align}
  \frac{\epsilon(\omega)}
       {\epsilon_0}
=
  \frac{\mu(\omega)}
       {\mu_0}
&=
  1
 -
  \frac{\omega_p^2}
       {\omega \left( \omega + \imath \gamma\right)}
.
\end{align}

\begin{figure}
\includegraphics[angle=-90,width=0.80\columnwidth]{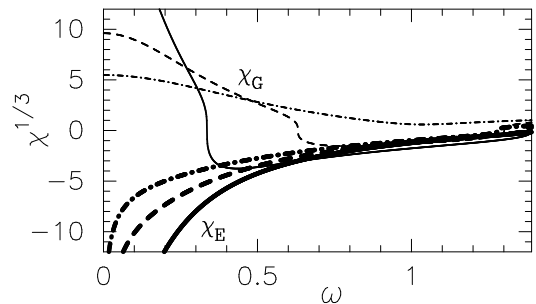}
\caption{
A double ($\epsilon$ and $\mu$)
 plasmon system exhibiting NRI, 
 with both plasma frequencies being $\omega_p \simeq 1.4$. 
It compares 
 the ($\chi_{E}$) NPVE condition (thick lines),
 to the ($\chi_{G}$) NPVG one (thin lines).
The results shown are for $\gamma=0.02$ (solid lines), 
 $\gamma=0.04$ (dashed lines), 
 and $\gamma=0.06$ (dot-dashed).
}
\label{fig-plasmon}
\end{figure}

A simple test to evaluate the presence of NPVG
 (and at the same time test our generalized causality-based criterion
 in Eq.~(\ref{eqn-integralRHS})),
 is to increase the losses
 whilst comparing it against the NPVE condition.
The results can be seen in Fig.~\ref{fig-plasmon},
 where $\chi^{1/3}$ is plotted to accommodate the vertical range.
For sufficiently weak losses
 ($\gamma \ll \omega$)
 the criteria agree, 
 with both the $\chi_{E}$ and $\chi_{G}$ curves remaining below zero.
However,
 as the losses get stronger,
 the $\chi_{G}$ and $\chi_{E}$ start to disagree.
Nevertheless,
 we can see that in the preferred region of $\omega \simeq 1$,
 where $\epsilon = \mu \simeq -1$,
 they disagree only for very large losses.
Here the $\chi_{G}$ criterion works relatively well
 because the plasmonic responses
 vary both smoothly and monotonically,
 hence $v_{g}$ does not change sign
   and remains in accord with $v_{E}$.

%

Our next example is again motivated by simplicity, 
 but also by the possibility of creating NRI in atomic gases.
In a gas, 
 it is possible to design pumping schemes
 that create gain \cite{Anant-AB-2007arXiv,Orth-EK-2007arXiv}, 
 but the freedom to manipulate the optical properties
 relies mainly on the dielectric response ($\epsilon$).
Here we consider two matched pairs of Lorentz resonances, 
 so that $\epsilon(\omega) = \mu(\omega)$, 
 and
~
\begin{align}
  \frac{\epsilon(\omega)}
       {\epsilon_0}
&=
  1
 +
  \frac{\sigma_1 \omega_1^2}
       {\omega_1^2 - \omega^2 - \imath \omega_1 \gamma_1}
 +
  \frac{\sigma_2 \omega_2^2}
       {\omega_2^2 - \omega^2 - \imath \omega_2 \gamma_2}
.
\label{eqn-doublelorentz}
\end{align}

%

\begin{figure}
\includegraphics[angle=-90,width=0.80\columnwidth]{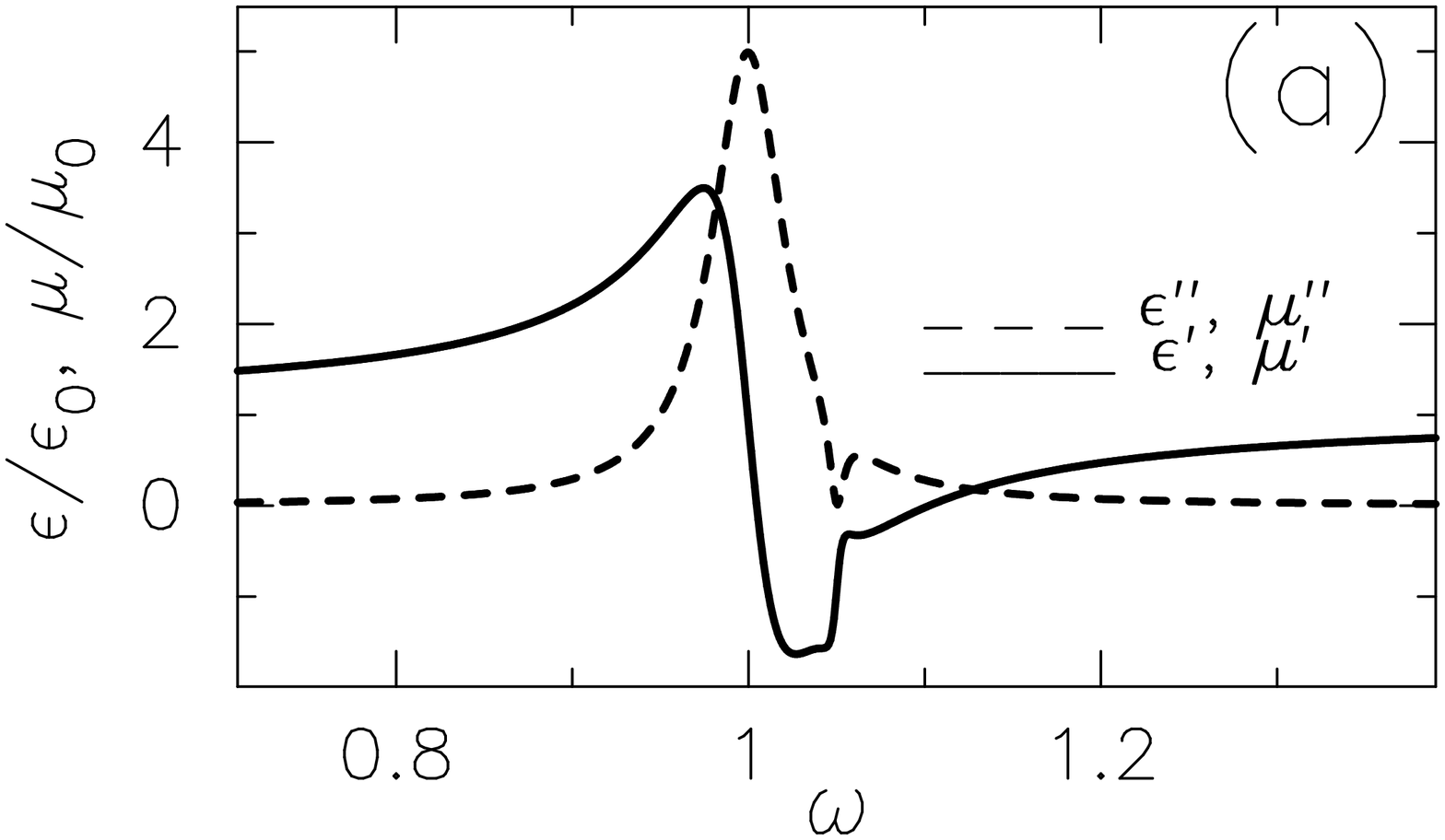}
\includegraphics[angle=-90,width=0.80\columnwidth]{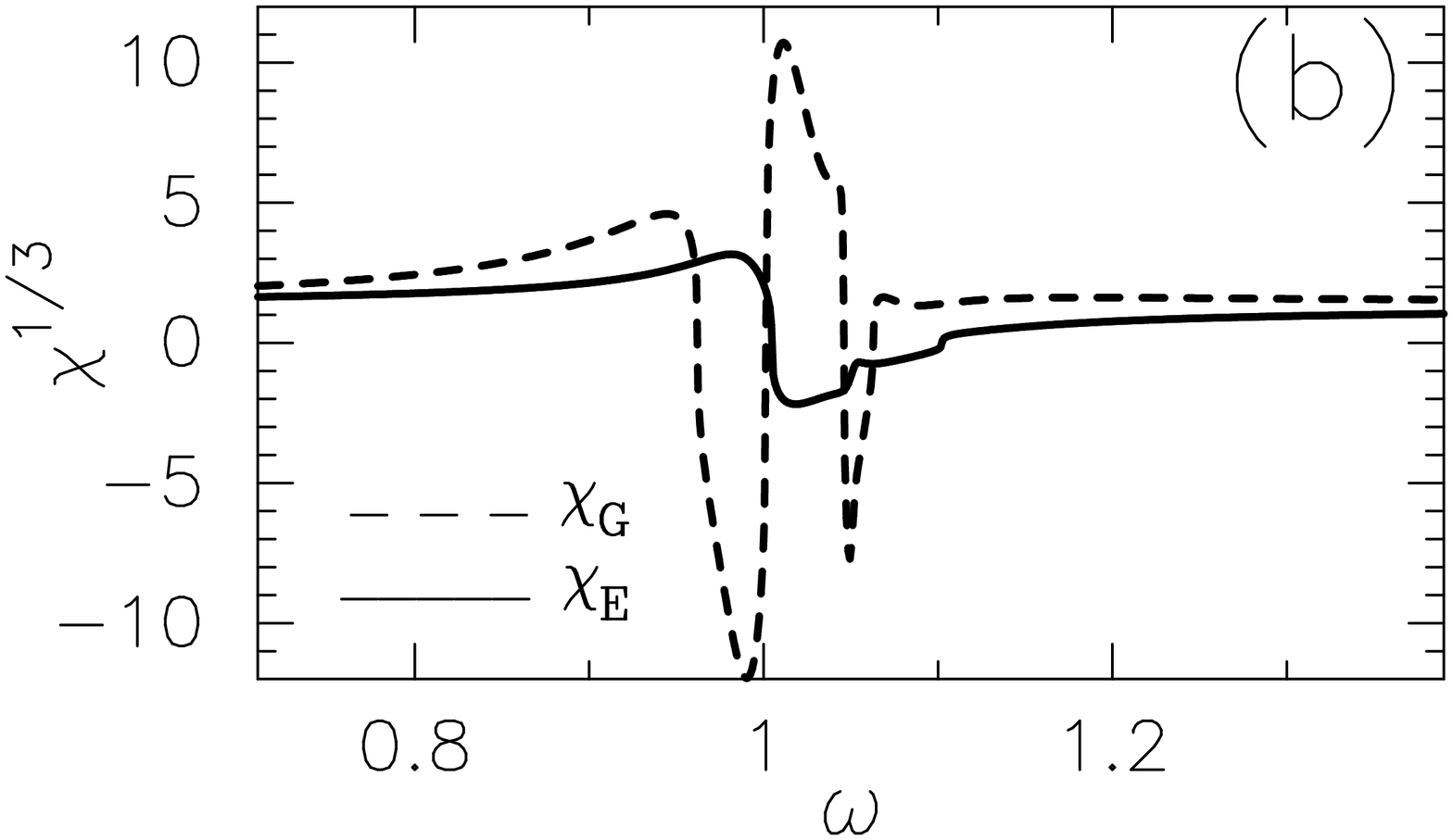}
\includegraphics[angle=-90,width=0.80\columnwidth]{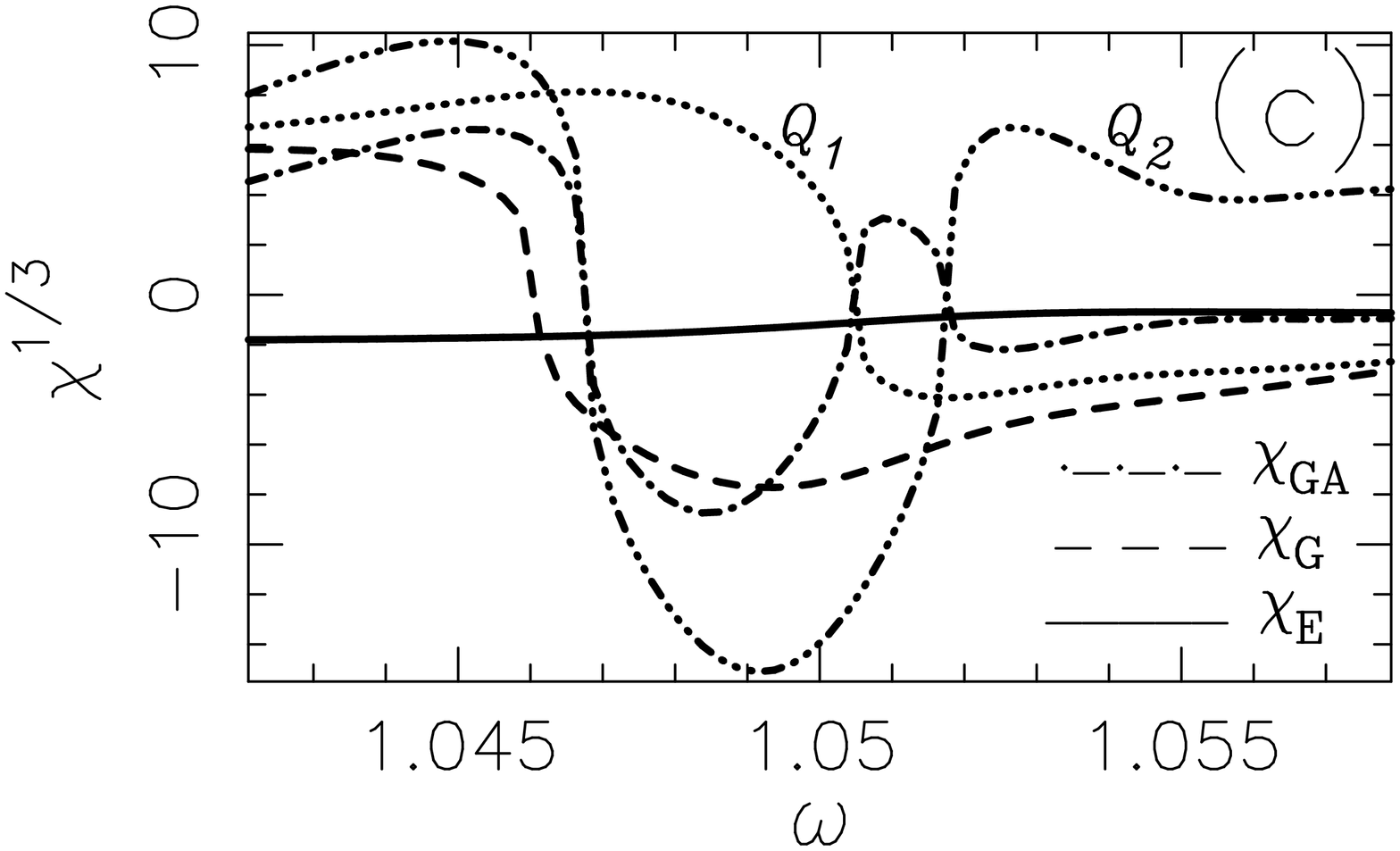}
\caption{
A system exhibiting narrowband NRI.
It combines a lossy resonance at $\omega_1=1$
 (with $\gamma_1=0.05, \sigma_1=-5$),
 and an active one at $\omega_2=1.05$
 (with $\gamma_2=0.01$ and $\sigma_2=1.02$).
(a)
 The real parts and imaginary parts of $\epsilon$ and $\mu$.
(b)
 Comparison of the NPVE ($\chi_{E}$) 
  and NPVG ($\chi_{G}$) criteria.
(c)
 Expanded view around $\omega_2=1.05$,
 showing also the NPVG approximation from Eq.~(\ref{eqn-integralRHSapprox})
 (labeled $\chi_{GA}$), 
 and $Q_1$ and $Q_2$.
} \label{fig-offset}
\end{figure}

We focus on a dominant lossy resonance $(\sigma_1 <0$),
 with a weaker, 
 offset, 
 \emph{active} resonance $(\sigma_2 > 0$)
 providing sufficient gain to induce near-transparency
 at a chosen observation frequency\footnote{Note 
  that this has a causal loss profile
  containing a minimum -- a situation supposedly excluded
  by the Eqs.~(5,6) and related discussion in \cite{Stockman-2007prl}.}.
$\epsilon$ and $\mu$
 are chosen equal apart from a scale factor $\epsilon_0/\mu_0$,
 and are shown on Fig.~\ref{fig-offset}(a), 
 where we see that near transparency
 has been achieved at the cost of increased dispersion,
 with $\epsilon'$ varying strongly where
 $\omega \simeq \omega_2$.
Note how the sign of $\chi_{G}$ 
 swaps back and forth according to the gradients
 of $\epsilon'$ and $\mu'$, 
 even though the values of $\epsilon'$ and $\mu'$ themselves
 change very little: 
 the utility of the $\chi_{G}$ criterion
 depends entirely on whether $v_{g}$ has the same sign
 as $v_{E}$ at the frequency of interest.

The narrowband region of low loss in this system
 makes it ideal for examining 
 our NPVG criterion of Eq.~(\ref{eqn-integralRHS})
 in more detail.
First, note that 
 there is an asymmetry about the loss minimum --
 below, 
 the two contributions to Eq.~(\ref{eqn-integralRHS}) reinforce
 to help satisfy the criterion; 
 above they partly cancel, 
 making NRI less likely; 
 this asymmetry is visible on Fig.~\ref{fig-offset}(b,c)
 around $\omega=1.05$.
At the minimum itself, 
 we can expect the ${Q}_1$ integral to be small
 since the integrand near $\omega$ will be not only small but odd; 
 the behaviour will then be dominated by that of ${Q}_2$ --
 and indeed on Fig.~\ref{fig-offset}(c) there is strong
 qualitative agreement between $Q_2$ and $\chi_{G}$.
The criterion therefore controls the width of allowed low-loss windows:
 a narrowband window will have a large ${Q}_2$,
 so that our criterion will be more easily satisfied.
This inference is related to Stockman's -- 
 it also demands sufficient loss close
 to the observation frequency,
 but does not require ${Q}_0={Q}_1=0$.

%

In conclusion, 
 we have derived a causality-based criterion for NRI 
 allowing for frequency dependent (dispersive) losses
 at the observation frequency.
Our new criterion is applicable to any medium with the linear response, 
 required by the Kramers-Kronig relations.
We investigated our causality-based criterion
 using some simple material response models,
 showing that since the group velocity $v_{g}$
 does not always match signs with the energy velocity $v_{E}$, 
 the NPVG and NPVE forms of NRI are not equivalent.
Since NPVE (i.e. $\vec{P} \cdot \vec{k} < 0$) is usually 
 the preferred condition for NRI, 
 this difference needs to be taken into account 
 before causality-based NRI conditions are utilized.
Nevertheless,
 our causality-based NPVG criterion provides unique insight 
 into how the global response of the material
 affects its local performance.

%
The authors acknowledge financial support from the EPSRC
 (EP/E031463/1, EP/G000964/1);
 and discussions with collaborators at 
 the University of Salford led by A.D. Boardman,
 and at the University of Surrey led by O. Hess.



%


%

\appendix

%
\section{Appendix: Previous commentary on Stockman result}

Stockman's criterion has been subject to recent comment 
 by Mackay and Lakhtakia \cite{Mackay-L-2007prl-c}, 
 who constructed a dispersion relation
 designed to provide a counterargument.
Stockman then disputed \cite{Stockman-2007prl-r} 
 the physical plausibility of this,
 and remarked that it did not minimize NPV losses very effectively.
While the acceptable level of losses varies according to the application, 
 the limitations of Stockman's result
 make it hard to determine how losses might be minimized 
 whilst still maintaining NPV over some frequency window.

%
\section{Appendix: Symmetric loss and gain}

\begin{figure}
\includegraphics[angle=-90,width=0.80\columnwidth]{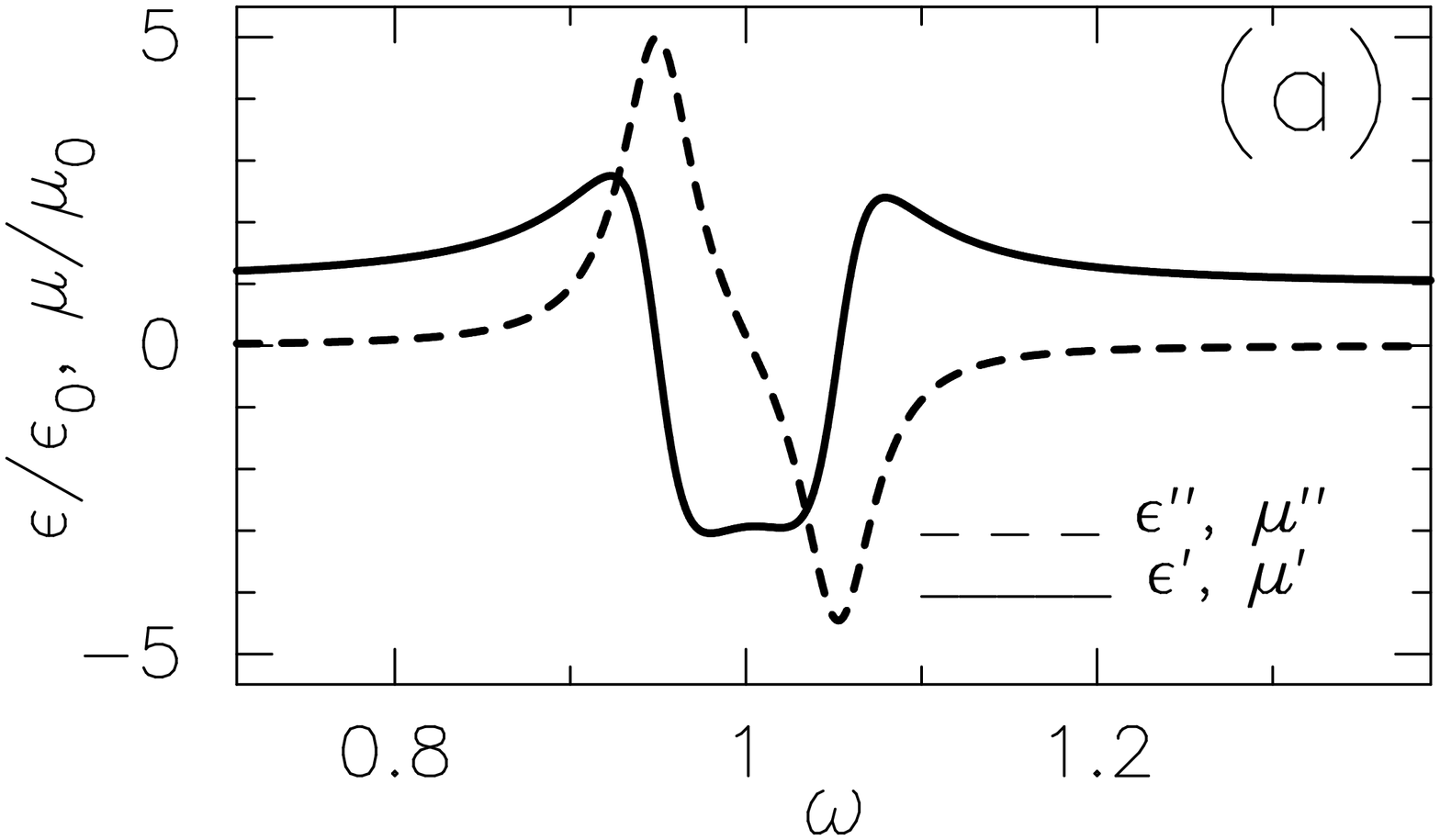}
\includegraphics[angle=-90,width=0.80\columnwidth]{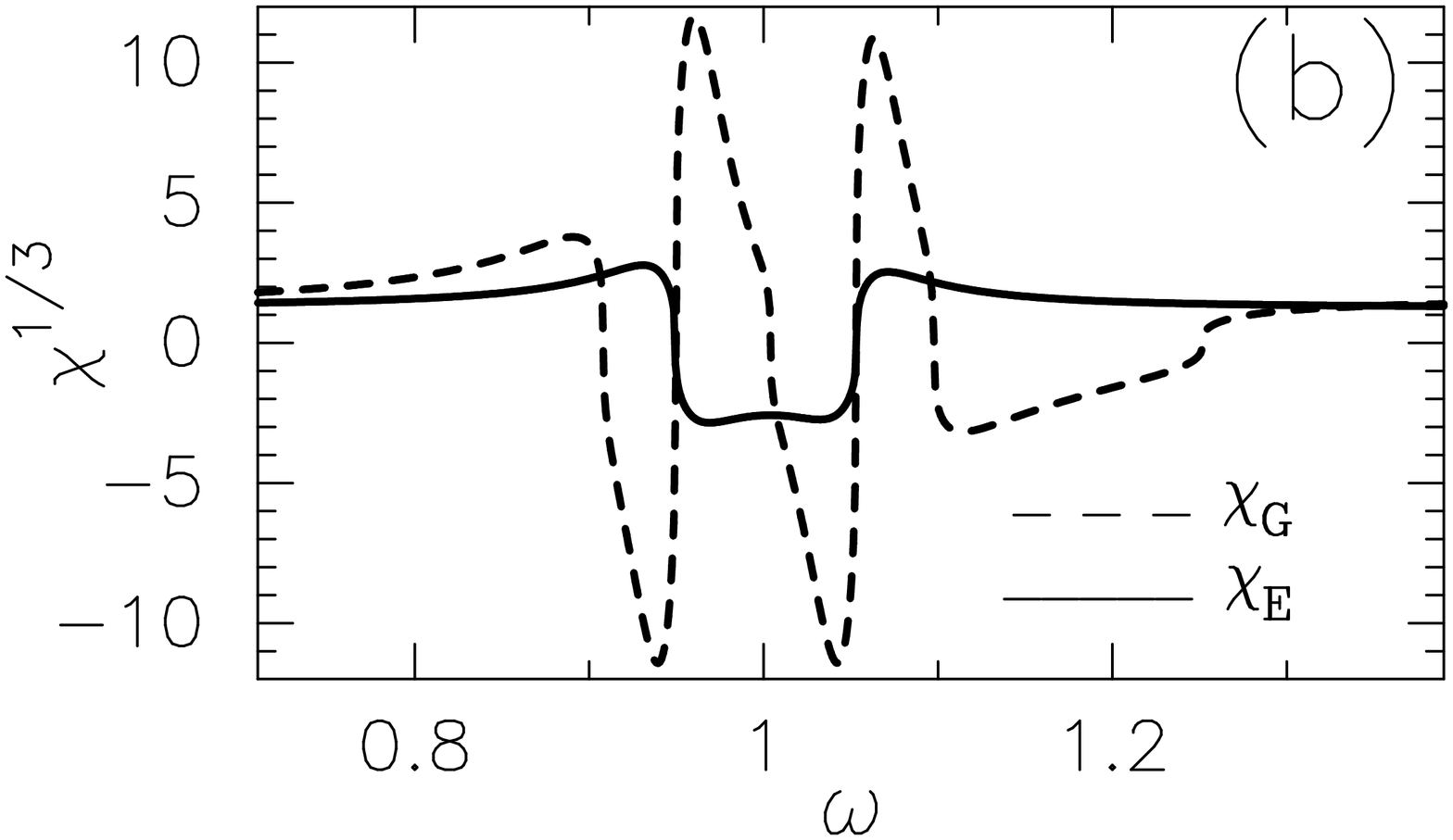}
\caption{
A system exhibiting a wideband and weakly dispersive 
 region of NRI (NPVE) near $\omega \simeq 1$.
It combines
 a lossy resonance at $\omega_1=0.95$, $\sigma_1=5$ and $\gamma=0.05$, 
 balanced against an active one at $\omega_2=1.05$,
 with $\sigma_2=-\sigma_1$ and $\gamma_2=\gamma_1$.
(a)
 The real parts and imaginary parts of $\epsilon$ and $\mu$.
(b)
 Comparison of the NPVE and NPVG conditions.
}
\label{fig-symmetric}
\end{figure}

Next
 we combine matched active and passive resonances 
 (i.e. $\sigma_1 = -\sigma_2$ in Eq.(\ref{eqn-doublelorentz}))
 offset either side of the observation frequency
 (see  Fig.~\ref{fig-symmetric}(a)),
 as in e.g. Anant et al. \cite{Anant-AB-2007arXiv}.
This configuration achieves a rather wide NPVE bandwidth -- 
 if it could be engineered,
 then it would be an attractive system for use.
However,
 just as for the offset-gain model above,
 the NPVG $\chi_{G}$ criterion performs poorly 
 compared to the NPVE one because of the way
 the sign of $v_{g}$ alternates while following the detailed behaviour
 of $\epsilon'$ and $\mu'$.

\end{document}